\documentclass[a4paper,11pt]{article}
\usepackage{pos}
\usepackage{physics}
\usepackage{subcaption}
\usepackage{wasysym}
\usepackage{xcolor}
\usepackage{booktabs}
\usepackage{multirow}

\makeatletter
\renewcommand{\tableofcontents}{
  \section*{Contents}
  \@starttoc{toc}
}
\makeatother

\title{Determination of $\alpha_S$ in the $SU(3)$ Yang-Mills theory}
\author*{Isabella Leone Zimmel}
\author{Alberto Ramos}
\affiliation{Instituto de Física Corpuscular (IFIC), CSIC-Universitat de València, 46071, Valencia, Spain}
\emailAdd{ilezim@ific.uv.es}
\emailAdd{alberto.ramos@ific.uv.es}

\abstract{The decoupling strategy allows one to obtain the value of the strong
coupling in QCD from the running in pure gauge. Here we present our
strategy to determine the running in the $SU(3)$ Yang-Mills theory. We
use a finite-volume scheme with twisted boundary conditions and a
step-scaling approach based on a gradient-flow coupling. We show
preliminary results for the continuum extrapolation of the step-scaling
function. Compared with other finite-volume approaches, we
expect a reduced statistical error and absence of linear cutoff
effects due to the translational invariance
of the boundary conditions.
}

\FullConference{The 42nd International Symposium on Lattice Field Theory (LATTICE2025)\\
 2-8 November  2025\\
Tata Institute of Fundamental Research, Mumbai, India\\}

\begin{document} 
\maketitle

\section{Introduction}

The strong coupling enters in all perturbative calculations of strong
processes, and its uncertainty contributes to a significant fraction
of the total theoretical uncertainty of several LHC cross-sections~\cite{Salam:2017qdl}.
The most precise determinations so far were achieved
on the lattice, with a precision of about $5 \permil$~\cite{Brida:2025gii}.

A key element of the strategy in~\cite{Brida:2025gii}
is the introduction of the decoupling method~\cite{FlavourLatticeAveragingGroupFLAG:2024oxs, Brida:2021xwa, DallaBrida:2019mqg}.
This approach uses the exact relation between the $\Lambda$ parameters of QCD and the $SU(3)$ Yang-Mills theory,
which exists as the latter is the heavy quark mass limit of the former.
This new strategy has put more emphasis on the pure gauge determination
of the $\Lambda$ parameter: no longer as a test-bed for new ideas, but as
part of a precision physics program in the determination of $\alpha_S$.
In fact, in~\cite{Brida:2025gii} about $20\%$ of the final error squared
in the QCD coupling originates in the pure gauge determination of $\Lambda$.
This work used as input the pure gauge determination of~\cite{DallaBrida:2019wur},
that quotes a total uncertainty of about $1.5\%$ in $\sqrt{t_0}\Lambda$.

In this contribution we present our strategy for an improved determination of the $SU(3)$ Yang-Mills theory’s $\Lambda$ parameter, with a sub-percent target precision. The key elements of this project are:
\begin{itemize}
\item{Finite-size scaling techniques~\cite{Luscher:1991wu} to avoid large systematics associated with the inversion of a perturbative series.} 
\item{A gradient flow finite-volume scheme with twisted boundary conditions~\cite{Ramos:2014kla, Luscher:2010iy} for improved precision.}
\item{A split of the computation of the step-scaling function in finite volume into a change of the renormalization scale followed by a change of the volume for reduced systematics related to the continuum extrapolation~\cite{Nada:2020jay}.}
\end{itemize}  
The rest of this contribution is dedicated to justifying
these choices using some preliminary simulations.

\section{Setup}
\subsection{Finite-size scaling}
The $\Lambda$ parameter is a RGI mass scale, given by:
\begin{eqnarray}
  \Lambda_{\rm x} &=& \mu\;\varphi_{\rm x}(\bar{g}_{\rm x}(\mu))\, \\
    \varphi_{\rm x}^{}(\bar g_{\rm x}) &=& ( b_0 \bar g_{\rm x}^2 )^{-b_1/(2b_0^2)} 
        e^{-1/(2b_0 \bar g_{\rm x}^2)} \times \exp\left\{-\int\limits_0^{\bar g_{\rm x}} {\rm d} g\ 
        \left[\frac{1}{\beta_{\rm x}(g)} 
          +\frac{1}{b_0g^3} - \frac{b_1}{b_0^2g} \right] \right\}\,,
      \label{eq:varphi}
\end{eqnarray}
where $\rm x$ represents the scheme. 

The challenge in determining $\Lambda$ resides in the evaluation of
the last integral, since it requires knowledge of the non-perturbative
$\beta$ function in the corresponding scheme.
In finite-size scaling methods this is extracted from the step-scaling function.
The step-scaling function quantifies the change in the running
coupling after a step in the renormalization scale by a factor $s$ (typically 2); it is thus a discrete version of the $\beta$ function

\begin{equation}
  \label{eq:ssf}
  \sigma_s(u) = \bar g^2\Big(\frac \mu s\Big)\Big\vert_{\bar g^2(\mu)=u}\,,
\end{equation}
and thanks to the exact relation
\begin{equation}
  \ln\Big(\frac {\mu_2}{\mu_1}\Big) = \int_{\bar g(\mu_1)}^{\bar g(\mu_2)} dx\; \frac 1{\beta(x)} \quad\Rightarrow\quad
  - \ln(s) = \int_{\sqrt{a}}^{\sqrt{b}} dx\; \frac 1{\beta(x)} \qquad\quad a=\textstyle \bar g^2,\;b=\sigma(\bar g^2),
\end{equation}
it allows the determination of pieces of
the difficult integral in Eq.~(\ref{eq:varphi}).
This quantity is well suited to lattice studies
in combination with finite-volume schemes, i.e.,
in which the box size is tied to the renormalization scale by the relation
$L\propto 1/\mu$, where a step by a factor $s$ in the renormalization scale is
achieved by a corresponding change in the lattice volume by the same factor.

\subsection{Choice of scheme}

In order to apply the idea of finite-size scaling one needs a
definition of a non-perturbative coupling in finite volume. 
This amounts not only to a choice of observable, but also to a choice
of boundary conditions, which in finite volume are relevant.

\subsubsection{Twisted boundary conditions}

Although it may seem that the most straightforward setup would consist
in imposing periodic boundary conditions on the fields in all four
Euclidean directions, it is well known that perturbation theory in
this setup is difficult due to the presence of zero-momentum
modes~\cite{Coste:1985mn}. For this reason most works using
finite-size scaling techniques typically use
Dirichlet boundary conditions (i.e. the Schr\"odinger
Functional~\cite{Luscher:1992an}). Here we use twisted boundary conditions as an
alternative, and the interested reader is invited to consult the
original references~\cite{GonzalezArroyo:1982ub,tHooft:1981nnx},
as well as~\cite{deDivitiis:1994yz,Ramos:2014kla},
in which the same version of twisted boundary conditions is used.
The main advantage of this scheme is that, due to the invariance under
translations, coupling definitions are free of linear cutoff effects.

\subsubsection{Gradient flow couplings}

As for the choice of observable we will use couplings derived from
the Gradient Flow (GF)~\cite{Luscher:2010iy}. 
The GF defines a diffusion-like process for the gauge field $A_\mu(x)$
by evolving it along the flow equation
\begin{equation}
  \label{eq:flow}
  \frac{d B_\mu(x,t)}{dt} = D_\nu G_{\nu\mu}(x,t)\,,\quad
  B_\mu(0,x) = A_\mu(x)\,.
\end{equation}
Note that the flow time has units of length squared.
Composite gauge invariant operators are automatically renormalized by
the flow. 
in particular the dimensionless combination of the flow time and action density
\begin{equation}
        \label{eq:et}
  t^2\langle E(x,t)\rangle = -\frac{t^2}{2} \langle{\rm
    Tr} G_{\mu\nu}(x,t)G_{\mu\nu}(x,t) \rangle\,,\quad
  \left(   G_{\nu\mu} = \partial_\nu B_\mu - 
  \partial_\nu B_\mu + [B_\nu,B_\mu] 
 \right)\,.
\end{equation}
is automatically renormalized~\cite{Luscher:2011bx}. 
Being a dimensionless quantity that depends on a scale $\mu =
1/\sqrt{8t}$, it is a convenient quantity for coupling definition.
In finite-size scaling applications the renormalization scale is
linked to the size of the system $\sqrt{8t}=cL$, (with $c\in
[0.2,0.5]$), and the coupling is defined via
\begin{equation}
  \bar g^2_{\rm TWBC}(\mu) = \mathcal N^{-1}(c)\, \frac {t^2\langle E(t)\; \delta_Q  \rangle}{\langle \delta_Q\rangle}\Big|_{\mu^{-1} = \sqrt{8t}=cL}\,.
\end{equation}
where the normalization $\mathcal N(c)$ is a known factor for our choice of
boundary conditions~\cite{Ramos:2014kla}, and $\delta_Q$ projects to
the sector of zero topology as a way to bypass the severe topology freezing
problem of our finite-volume setup~\cite{Fritzsch:2013yxa,Bonanno:2024nba}.

\subsection{Strategy}

The computation of the step-scaling function Eq.~(\ref{eq:ssf}) from
lattice simulations is in principle straightforward. 
One simulates at a value of the bare coupling $g_0$ on a lattice of
size $L/a$.
Keeping the bare coupling fixed (and hence the lattice spacing $a$),
one doubles\footnote{Although many scale factors $s$ are possible, here
we use $s=2$.} the number of points in the lattice simulation $2L/a$ and
gets a lattice approximation of the step-scaling function $\Sigma(u,L/a)$. 
Simulating several pairs of lattices allows to perform a continuum extrapolation
\begin{equation}
  \sigma(u) = \lim_{a/L\to 0} \Sigma(u,L/a)\,.
\end{equation}
This basic strategy has long been the standard way to determine the
step-scaling function, especially when the Schrödinger-functional
coupling is used~\cite{Luscher:1992an}, because in that approach the
renormalization scale is fixed by $L$.
For gradient-flow couplings, by contrast, the renormalization scale is
set by the flow time $t$; although $t$ can be related to the
spatial size $L$, as in finite-size scaling, the independence of the
two scales opens the door to alternative procedures.

In~\cite{Nada:2020jay} it was proposed to exploit this freedom to decompose the determination of the step-scaling function.
First, one measures the change in the coupling when the renormalization scale is lowered by a factor $s$ at fixed volume.
Second, the volume is increased by the same factor at fixed renormalization scale.
Thus, for $s=2$,

\[
\mathcal{J}_1(u)=g^2(\mu/2,L)\big|_{g^2(\mu,L)=u}\,,\qquad
\mathcal{J}_2(u)=g^2(\mu,2L)\big|_{g^2(\mu,L)=u}\,,
\]
and the step-scaling function is exactly
\[
\sigma(u)=\mathcal{J}_2\!\bigl(\mathcal{J}_1(u)\bigr)\,.
\]

Reference~\cite{Nada:2020jay} observed that most cutoff effects arise
from the change of scale ({\em i.e.}\ in the determination of $\mathcal{J}_1$), whereas
$\mathcal{J}_2$ involves only a change of volume and is therefore much
less affected by lattice artifacts.
On the other hand, the determination of $\mathcal J_1$ does not require
doubling the lattice volume, and therefore for a given dataset much more
data is available to perform this extrapolation.
Below we compare the direct determination of $\sigma(u)$ with the
two-step procedure and demonstrate the resulting reduction of
systematic uncertainties.

\section{Preliminary results}

\subsection{Numerical setup}

For the simulation we use the standard Wilson plaquette gauge action
\begin{equation}
  \label{eq:Waction}
  S_{\rm W}[U] = \frac{\beta}{6}\sum_{p}w(p)\,{\rm tr}(1-U_p)\,,
\end{equation}
where the sum extends over all plaquettes of the lattice, and $U_p$ denotes the ordered product of gauge links around plaquette $p$.  
The prefactor is $\beta=6/g_0^2$, with $g_0$ the \emph{bare} gauge coupling.  
Every weight $w(p)$ equals~1 except for one plaquette in each $(x,y)$ plane, where $w(p)=\mathrm{e}^{2\pi\mathrm{i}/3}$.  
Although this choice appears to break translational invariance, it in fact imposes twisted boundary conditions; see~\cite{Ramos:2014kla} for details.

For the simulations we employ the open-source package
$\texttt{LatticeGPU.jl}$.  
We alternate one hybrid Monte-Carlo update with $2L/a$ over-relaxation
sweeps; the latter is responsible for decorrelating measurements while
the former---simple to implement on a GPU---guarantees ergodicity.

To define our coupling we first have to use a lattice version of the
GF equation~(\ref{eq:flow}). 
We use the $\mathcal O(a^2)$ improved \emph{Zeuthen} flow
equation~\cite{Ramos:2015baa}
\begin{equation}
  \label{eq:impflowlat}
    a^2\left(\partial_t V_\mu(t,x)\right) V_\mu(t,x)^\dagger  =
    -g_0^2 \left(1 + 
      \frac{a^2}{12}\nabla_\mu^\ast\nabla_\mu^{}
    \right) \partial_{x,\mu} S_{\rm LW}[V]\,. \quad
    V_\mu(0,x) = U_\mu(x)  \,.
\end{equation}
This ensures that integrating the flow equations will not produce any
$\mathcal O(a^2)$ effects. 
One also has to choose a discretization of the energy density
observable $E(t)$ Eq.~(\ref{eq:et}). 
We again choose the $\mathcal O(a^2)$ improved combination of
plaquette and clover~\cite{Ramos:2015baa}
\begin{equation}
  \hat E(t) = \frac{4E_{\rm pl}(t)-E_{\rm cl}(t)}{3}.
\end{equation}
In order to project the coupling onto the trivial topological
sector we use the field-theoretical definition of the topological
charge $Q$ using the clover definition of the field strength. 
The projector is defined by
\begin{equation}
\delta_Q = \begin{aligned}\begin{cases}0\qquad &|Q| > 0.5\\ 1& |Q|\leq 0.5\end{cases}\end{aligned}
\end{equation}
Finally the coupling is defined as
\begin{equation}
  \label{eq:glatt}
  \bar g^2_{TWBC}(\mu, L/a) = \hat{\mathcal N}^{-1}(c, L/a) \frac {t^2\langle \hat E(t)\; \delta_Q  \rangle}{\langle \delta_Q\rangle}\Big|_{\mu^{-1}=\sqrt{8t}=cL},
\end{equation}
where the normalization $\hat{\mathcal N}(c,L/a)$ is computed in
lattice perturbation theory to ensure that, at tree level, the coupling
Eq.~(\ref{eq:glatt}) is free of lattice artifacts~\cite{Ramos:2014kla}. 

For this analysis we use lattice volumes $L/a =
8,10,12,16,20,24,32,40,48$ and bare couplings in the range $\beta \in
[5.9,15.0]$, and take measurements at $c = 0.2, 0.3, 0.4$.

\subsection{Data analysis and estimation of systematics}

Error analysis was performed with the package
ADerrors.jl~\footnote{Freely available at
  \url{https://igit.ific.uv.es/alramos/aderrors.jl}}. In this software
autocorrelations are estimated with the
$\Gamma$ method~\cite{Wolff:2003sm}, and errors are propagated using
automatic differentiation~\cite{Ramos:2018vgu}.

We compare the potential systematics of a direct determination
of the step-scaling with $c=\sqrt{8t}/L = 0.3$ with the split
determination using $c=\sqrt{8t}/L = 0.2$. The key systematics
to investigate is the continuum extrapolation. 
For this purpose it is convenient to look at the criteria in the FLAG review~\cite{FlavourLatticeAveragingGroupFLAG:2024oxs}. 
FLAG uses two main quantities as continuum extrapolation quality criteria. 
First the \emph{level arm} of the extrapolation, quantified by $\eta =
[a_{\rm max}/a_{\rm min}]^2$. 
Second the \emph{size of the extrapolation}, defined as the difference
between the extrapolated result of a quantity ($Q(0)$) and its value
at \(a_{\rm min}\) ($Q(a_{\rm  min})$), expressed in units
of the quoted uncertainty of the extrapolated quantity ($\sigma_Q$):
\begin{equation}
  \delta = \frac{Q(a_{\rm min}) - Q(0)}{\sigma_{Q}}\,.
\end{equation}
For the value at $a_{\rm min}$ it is more convenient to use the
prediction from the fitting functional form, in order to avoid the
(expected) fluctuations of the last data point.
We will explore the values of $\eta,\delta$ in the extrapolations that
enter our preliminary dataset.

\subsection{Direct determination of the step-scaling function}

\begin{figure}[h]
  \subcaptionbox*{ Measurements of the coupling at $L/a = 12$ and $L/a = 24$ at $c=0.3$. The combination $1/{\bar g^2} - 1/g_0^2$ is interpolated with a parabola. To extract $\Sigma(u = 2.8205, L/a = 12)$, the coupling in the largest lattice volume is evaluated at $g_0^2$ such that $g^2(g_0^2, L/a=12 ) = 2.8205$ %
}{\includegraphics[height=0.35\linewidth]{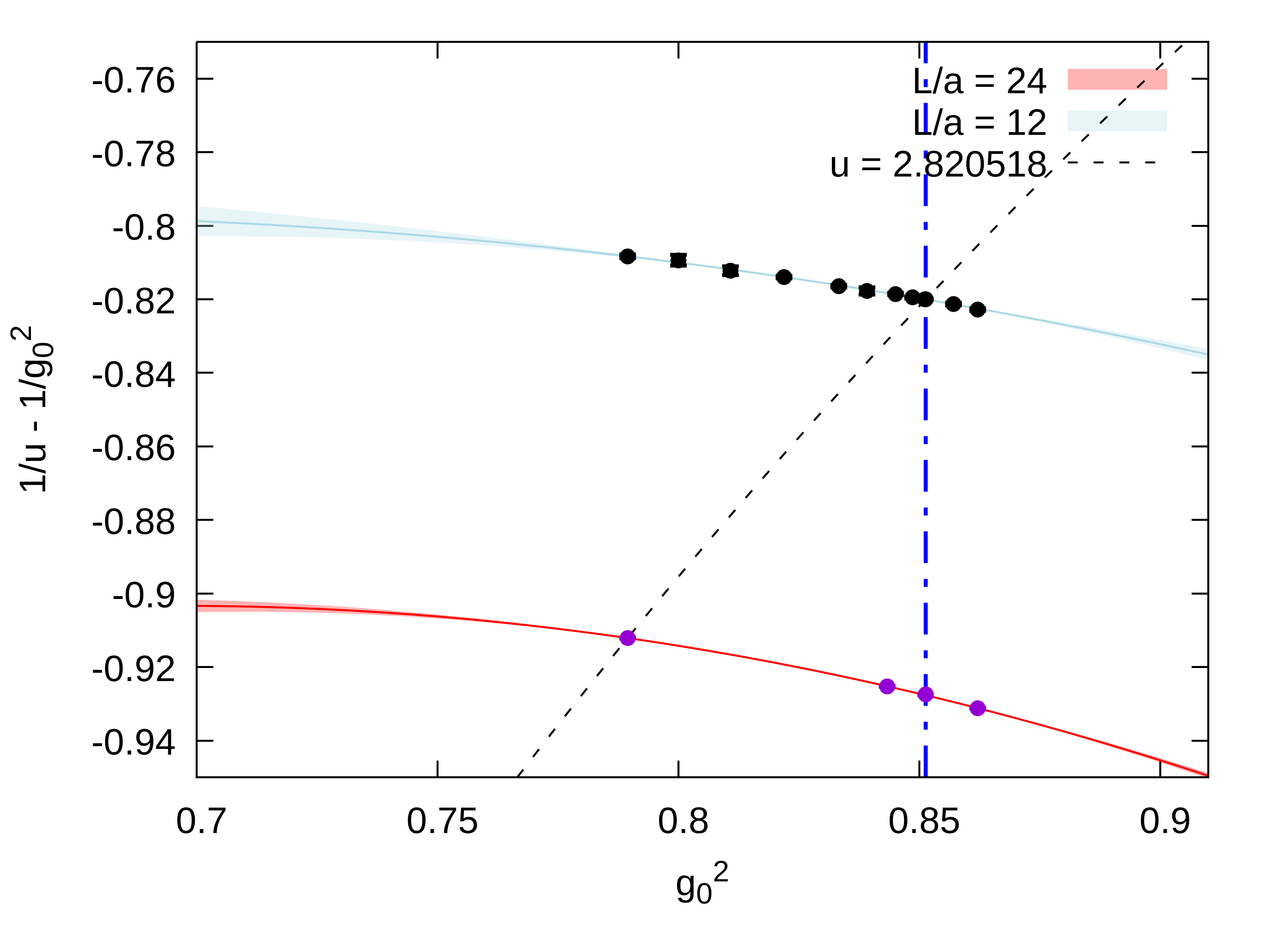}}\hfill%
\subcaptionbox*{Measurements of $J_1(u,L/a = 32)$ interpolated with a parabola to extract $J_1(u=3.2555,\;L/a=32)$. 
}{\includegraphics[height=0.35\linewidth]{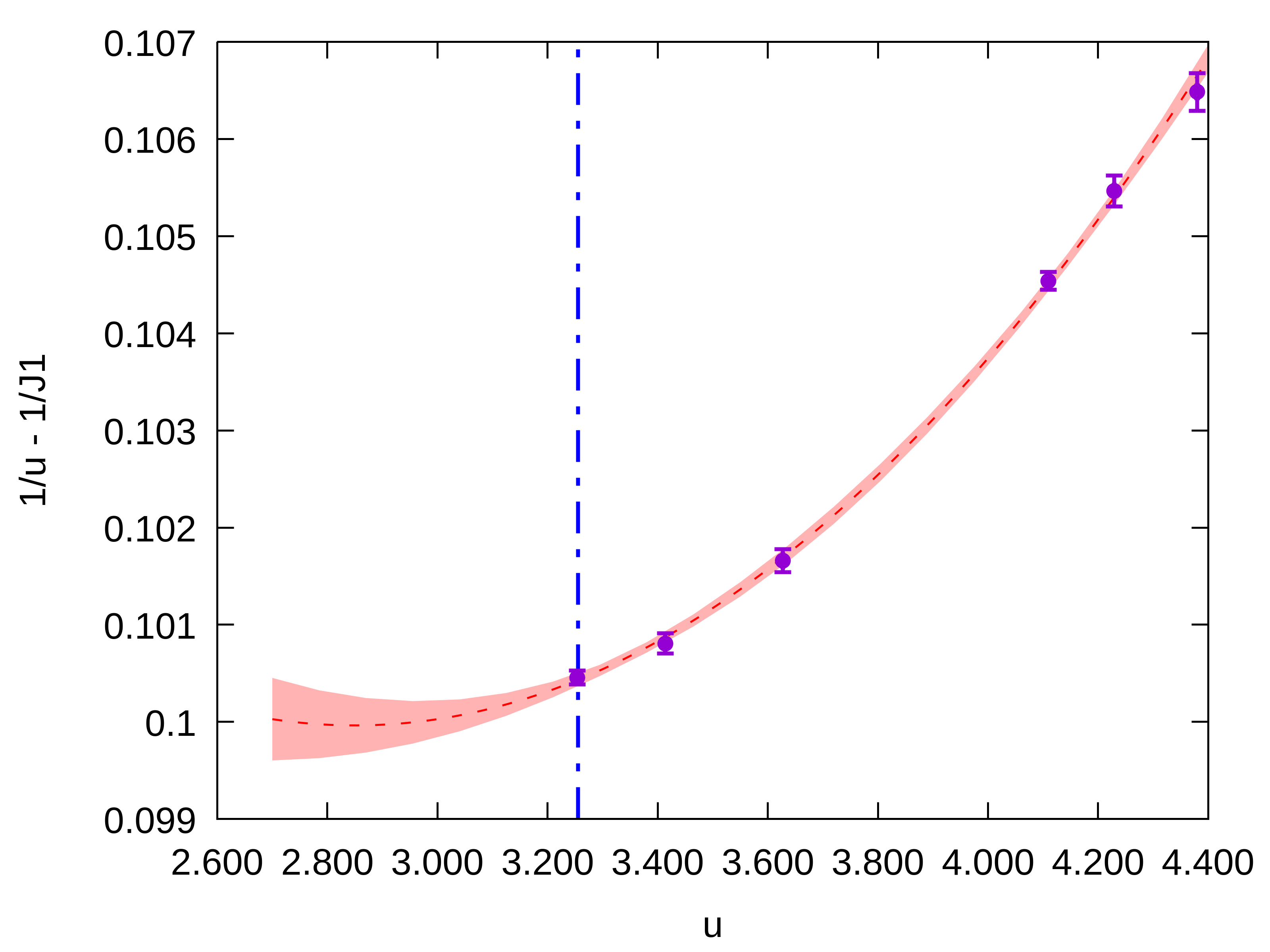}}\hfill%
  \caption{\label{fig::fitS_1} Tuning process to extract lines of constant physics for the continuum extrapolations of $\Sigma$ and $J_1$.}
\end{figure}

We consider the data at $c=0.3$, $L/a\in[8,24]$ and the corresponding $2L/a$.
We define a line of constant physics (LCP) by the condition $g_{\rm
  TWBC}^2(\mu,L/a) = u_t$,
and determine the values of $g_0^2=6/\beta$ that obey this LCP for
lattices of size $8, \ldots, 24$. The value $u_t= 2.8205$ is
given by the simulation at our largest lattice.
The tuning is better performed by fitting the combination
$1/{\bar g^2}-1/g_0^2$ to a polynomial, to remove the leading $g_0^2$ dependence in the coupling (see Figure~\ref{fig::fitS_1}).

Once the values of $\beta$ that correspond to this LCP have been
fixed, we determine the lattice step-scaling function
$\Sigma(u_t,L/a)$ by performing simulations at that value of the bare
coupling in a lattice of double the size.

The determination of the step-scaling function then amounts to
performing the continuum extrapolation.
We observe that all lattices $L/a > 8$ can be fitted
with a simple quadratic form. 
The result of this extrapolation (see Figure~\ref{fig::continuum_extrapolation_1}) has
$\eta= 2.25$ and $\delta = 2.27$.

\subsection{Splitting the determination of the step-scaling function}

\begin{figure}[h]
\subcaptionbox*{ Extrapolation of $\Sigma$.
}{\includegraphics[height=0.35\linewidth]{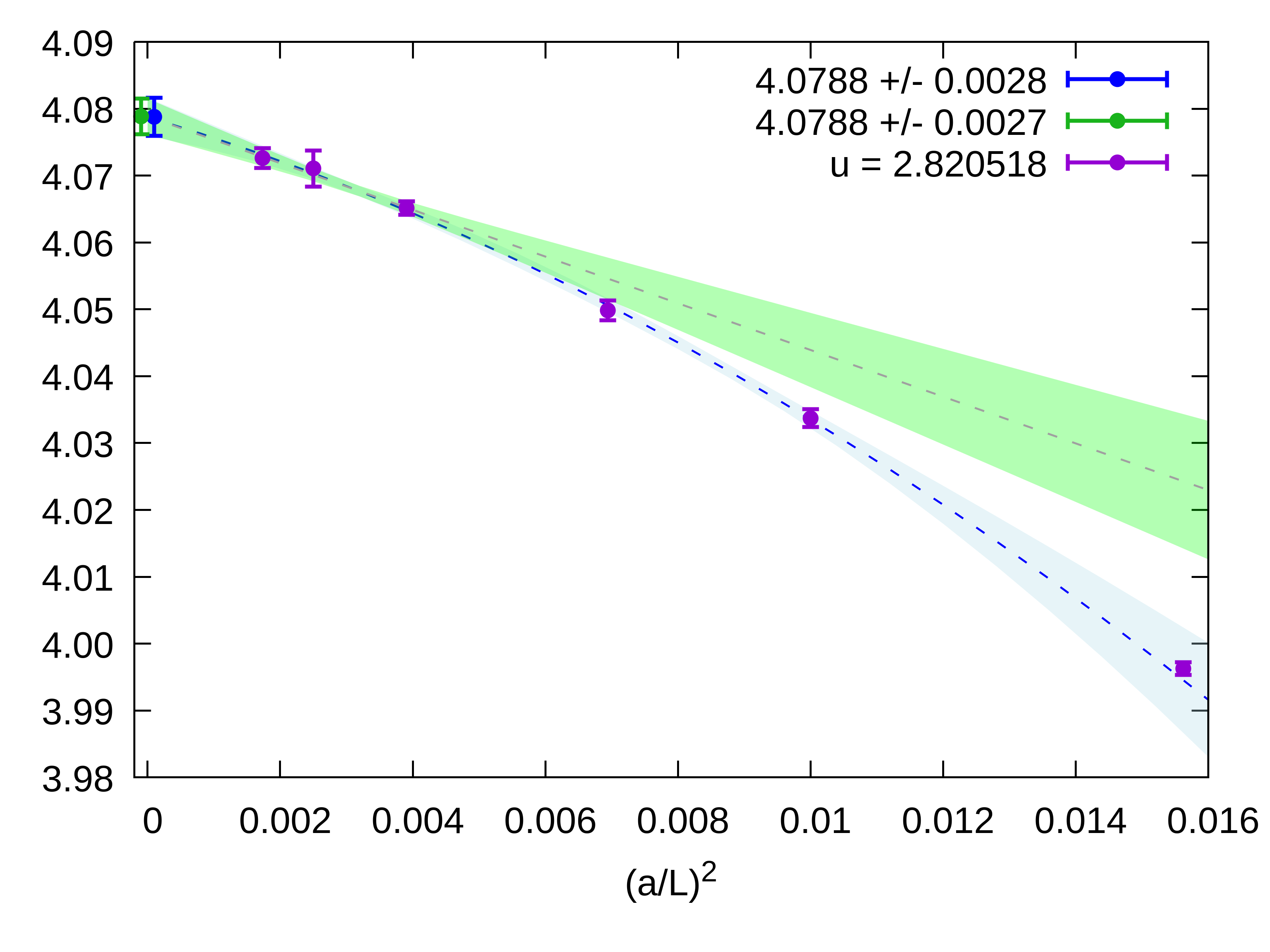}}\hfill%
\subcaptionbox*{ Extrapolation of $J_1$.
}{\includegraphics[height=0.35\linewidth]{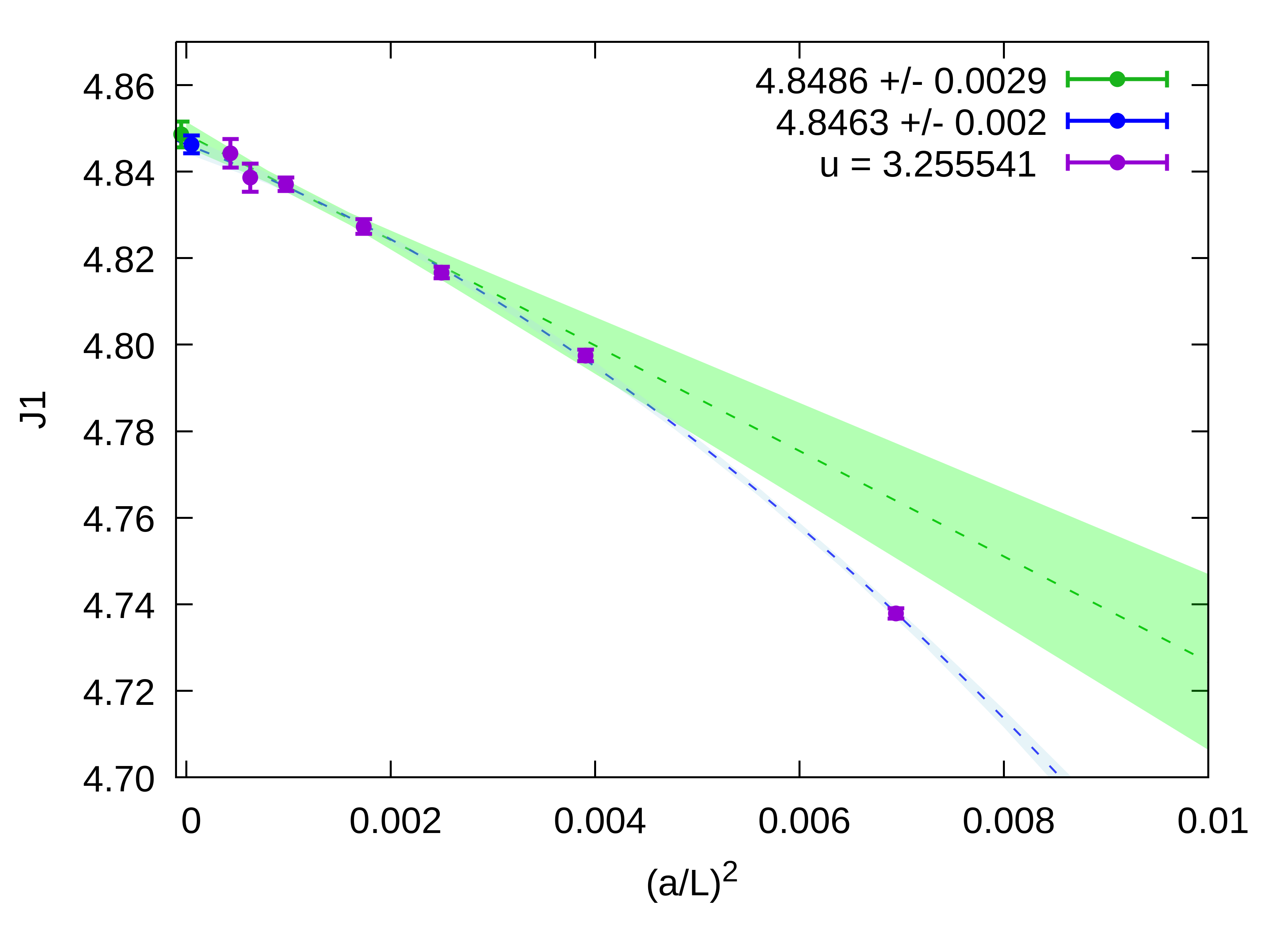}}\hfill%
\caption{\label{fig::continuum_extrapolation_1} Preliminary continuum extrapolations for $\Sigma$ and $J_1$, tuning at the central value of the coupling at $L/a = 48$, $\beta = 7.6$ at $c=0.3$ and $c=0.2$ respectively. A linear (quadratic) fit is shown in blue (green), in the range $L/a \geq 8$ ($L/a \geq 16$) for $\Sigma$ and $L/a \geq 12$ ($L/a \geq 20$) for $J_1$. Although both extrapolations are good, in the case of $J_1$ the arm of the extrapolation is shorter and more points are available in the linear scaling region (see Table~\ref{tab::quality_extrap}). }
\end{figure}

We used the available data at $c=0.2$ and $c=0.4$, $L\in[12,48]$, and tuned $u$ to the central value of the coupling obtained in the largest lattice at $L/a=48$, $\beta=7.6$. In the smaller lattices we interpolated $J_1(u, L/a)$ to extract $J_1(u_t)$. To subtract the leading $g_0^2$ dependence in $J_1$, we fitted the combination $1/u\;-\;1/J_1(u,L/a)$. For $L/a=40$, a single measurement compatible with the value at $L/a=48$ was available. This measurement was included after shifting the central value according to
\begin{equation}
  \frac{\Delta J_1(u,L/a)}{J_1^2(u,L/a)} = \frac{\Delta u}{u^2}.
\end{equation}
We then performed a linear and a quadratic extrapolation in $(a/L)^2$ of $J_1(u_t,L/a)$, and found the two results in agreement (see Figure~\ref{fig::continuum_extrapolation_1}).

\begin{figure}[h]
  \subcaptionbox*{Measurements of the coupling at $L/a=32;\; c=0.2$ and $L/a=16;\;c=0.4$ and the respective quadratic interpolation. The extracted value of $J_2^{-1}(u=3.2555, \;L/a=32)$ is shown in magenta. %
  }{\includegraphics[height=0.35\linewidth]{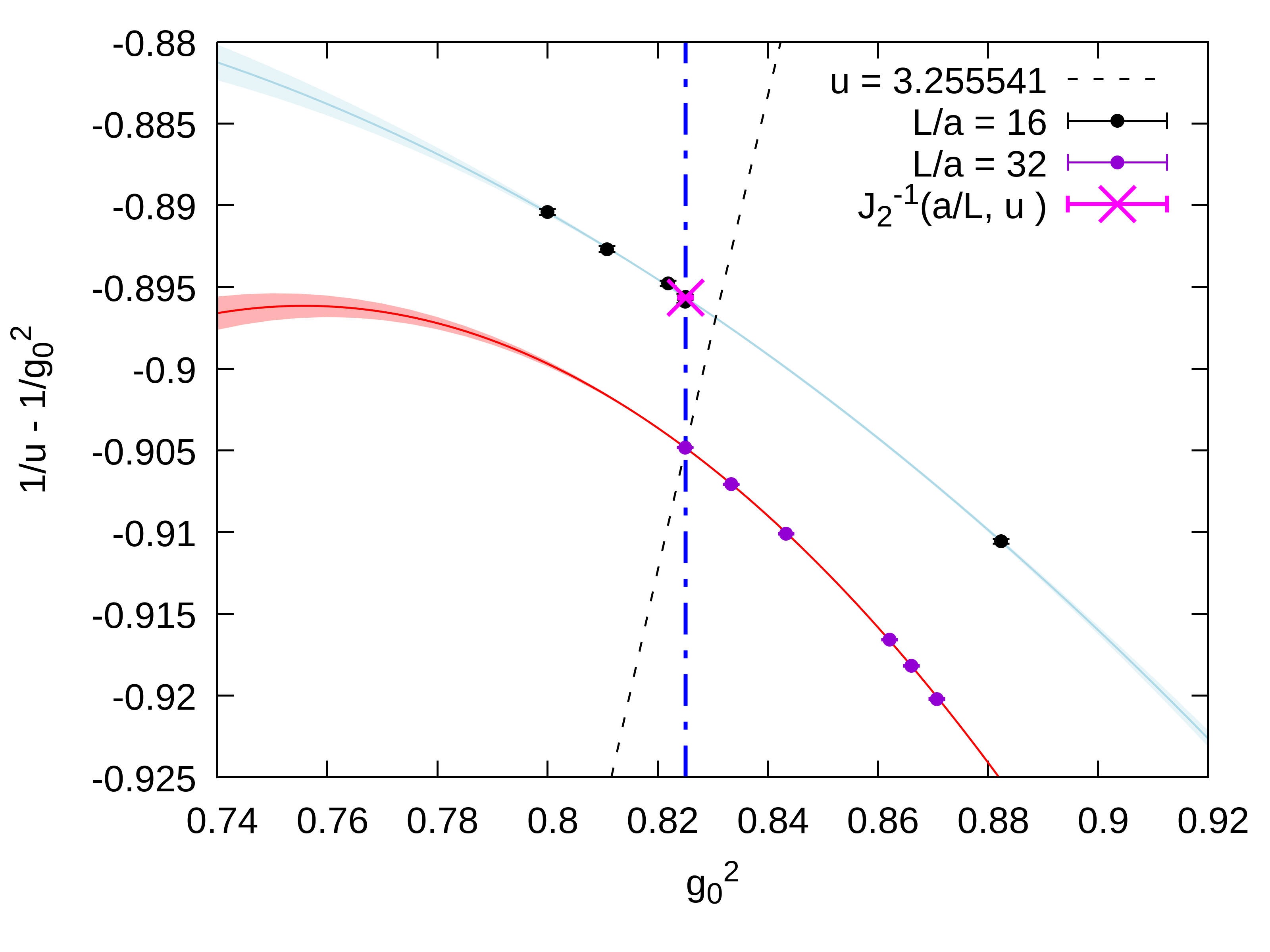}}\hfill%
\subcaptionbox*{Preliminary continuum extrapolation of $J_2^{-1}$. A linear extrapolation in $(a/L)^2$ of the available data is shown, and the result is shown in blue. As predicted the scaling violations in $J_2$ are very small.
}{\includegraphics[height=0.35\linewidth]{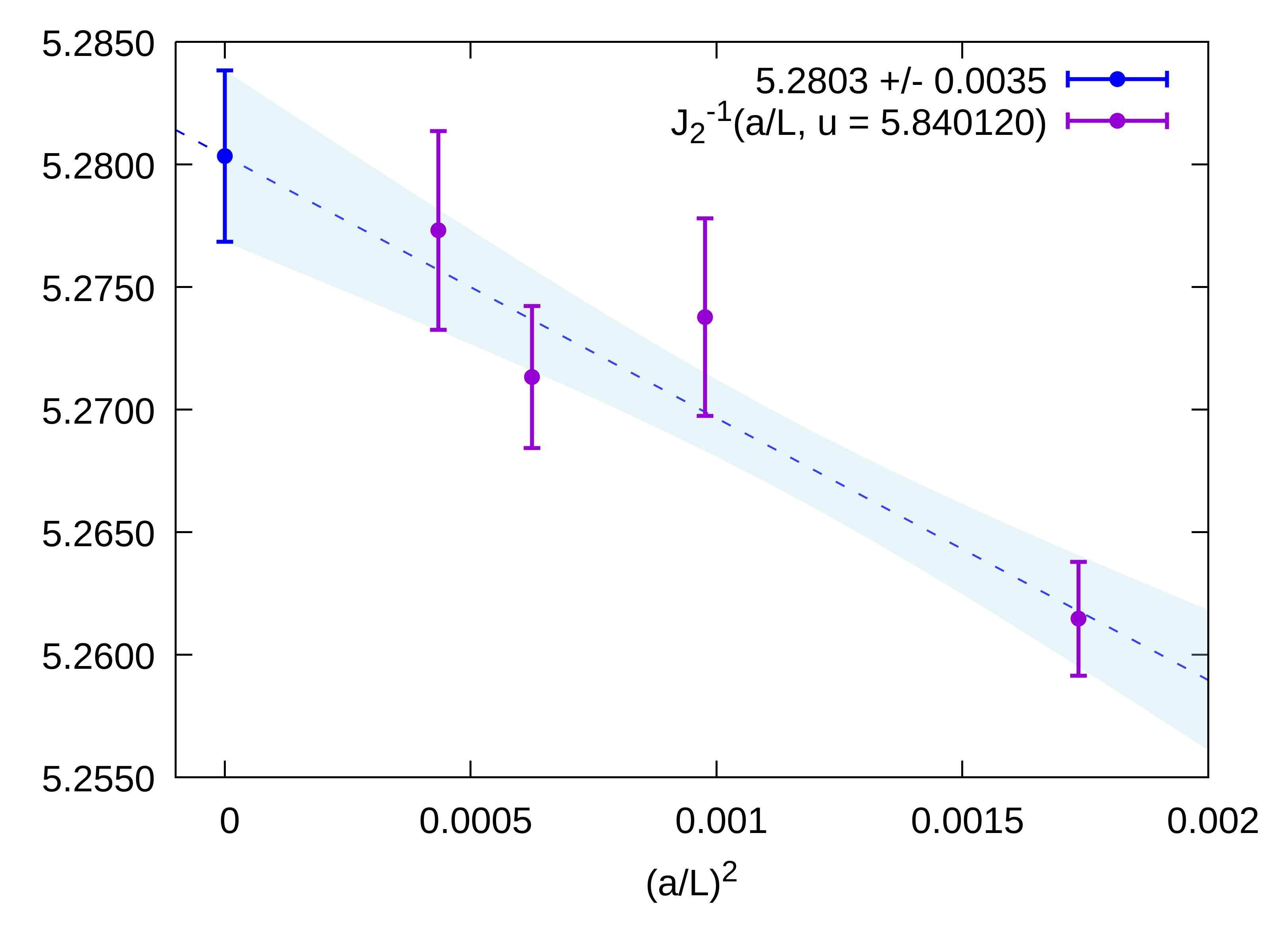}}\hfill%
  \caption{Tuning process and continuum extrapolation for the extraction of $\mathcal J_2$. \label{fig::fitJ2}}
\end{figure}

For the determination of $\mathcal J_2$ we followed a very similar procedure as for $\Sigma$, except that measures of the coupling at different values of $c = 0.2, 0.4$ for the two lattice volumes in each pair are used. In this case we extracted the inverse of the target function

\begin{equation}
  J_2^{-1}(u,L/a) = \bar g^2(g_0^2, L/2a, c)\Big\vert_{\bar g^2(g_0^2, L/a, 2c)=u}
\end{equation}
where we used the lattice volumes $L/a \in [20, 48]$ and the corresponding half-size lattices (see Figure~\ref{fig::fitJ2}); finally we perform a linear continuum extrapolation in $(a/L)^2$.

\begin{table}[h]
   \centering
   \begin{tabular}{cccccr}
     $Q\quad$ &$\delta$ &$\eta$& $\;N_{\text{points}}\;$&$\Delta Q / Q$ &$\hspace{7em}$ \\[0.5em]\toprule
     \multirow{1}{2em}{$\Sigma$}
 &2.27 &2.25 &3 &6.6e-04 &$u_t = $2.8205\\
\vspace{-0.7em}\\ 
     \multirow{2}{2em}{$J_1$}
 &1.80 &4.00 &4 &6.1e-04 &$u_t = $3.2555\\
 &1.33 &5.76 &5 &8.6e-04 &$u_t = $5.8401\\
\vspace{-0.7em}\\ 
     \multirow{2}{2em}{$J_2$}
 &0.35 &4.00 &3 &4.8e-04 &$u_t = $3.2555\\
 &1.33 &4.00 &4 &6.6e-04 &$u_t = $5.8401\\
\vspace{-1.5em}
   \end{tabular}\caption{The parameters $\eta =
     [a^2_{\text{max}}/a^2_{\text{min}}]$ (level arm) and $\delta$
     (size) are used to quantify the systematics in the continuum extrapolation. 
     In the FLAG review an extrapolation is considered good if $\eta >
     2$ and $\delta < 3$.
     The data shows that the split combination
     shows smaller systematics. The fifth column displays the relative error
     of the extrapolated value. \label{tab::quality_extrap}} 
\end{table}
The values of $\eta$, $\delta$ for all linear continuum extrapolations
for which a measurement at $L/a = 48$ was available, are shown in
Table~\ref{tab::quality_extrap}. We confirm our expectations: for the
same dataset, the systematics in the extraction of $J_1$ and $J_2$ is smaller: the size of the extrapolation is smaller, the level arm
larger, and more data is available in the linear $\mathcal O(a^2)$ regime. 
At the same
time, the precision of the extrapolated value is retained, as appears
from the relative errors shown in Table~\ref{tab::quality_extrap}.  

\begin{figure}[h]
\subcaptionbox*{ Extrapolations of $J_1$. The result of linear(quadratic) extrapolations is shown in blue(red). %
}{\includegraphics[height=0.32\linewidth]{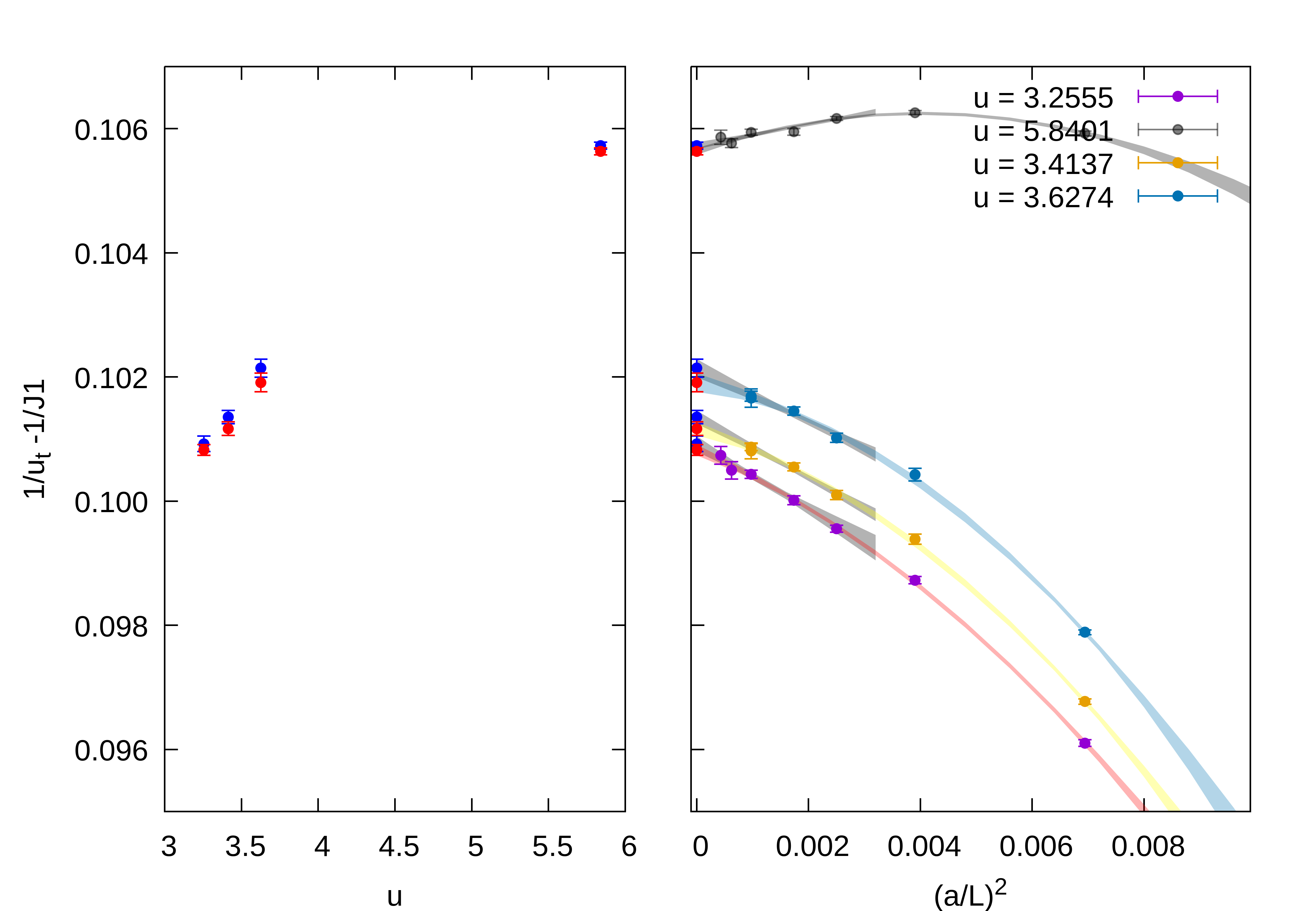}}\hfill%
\subcaptionbox*{ Extrapolations of $(J_2)^{-1}$. The result of linear extrapolations is shown in blue. %
}{\includegraphics[height=0.32\linewidth]{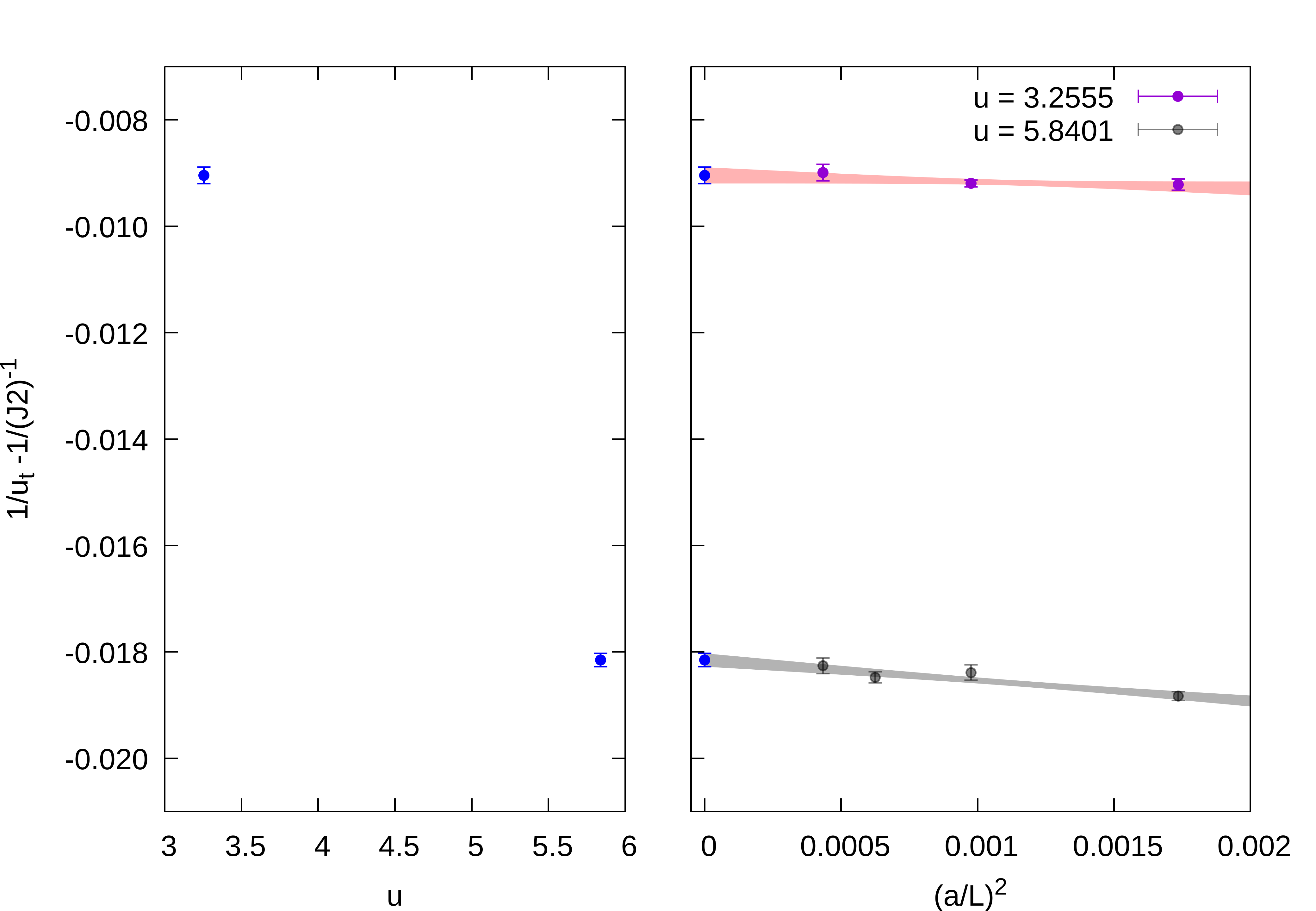}}\hfill%

\caption{\label{fig::continuum_extrapolation_2} Preliminary continuum extrapolations of $J_1$ and $J_2^{-1}$. To make errors more visible the combination $1/{u_t}-1/J_i$, with $J_i = J_1, J_2^{-1}$ is plotted, where $u_t$ fixes the line of constant physics.} 
\end{figure}

\vspace{-0.7em}\section{Conclusions}

\vspace{-0.5em} In our preliminary study we have investigated the systematics associated with the continuum extrapolation
of the step-scaling function. In particular we considered the possibility, suggested in \cite{Nada:2020jay},
to split the step-scaling function into a step in the renormalization scale followed by a step
in the volume and to extrapolate the two individually. Comparing this with the direct approach, we found better control on the systematics with the new method,
in which both extrapolations are of overall better quality in terms of the parameters $\delta$, $\eta$, due to a longer level arm and a smaller size of the extrapolation.


\acknowledgments
\vspace{-0.5em}
\noindent This work is partially supported by the Spanish Research Agency (Agencia Estatal de Investigación) through the grant PID2023-148162NB-C21. 

Numerical calculations have been performed on Artemisa, funded by the European Union ERDF and Comunitat Valenciana, and on the supercomputing infrastructure of the Galician Supercomputing Center (CESGA). The supercomputer FinisTerrae III have been funded by the NextGeneration EU 2021 Recovery, Transformation and Resilience Plan, ICT2021-006904, and also from the Pluriregional Operational Programme of Spain 2014-2020 of the European Regional Development Fund (ERDF), ICTS-2019-02-CESGA-3, and from the State Plan for Scientific and Technical Research and Innovation 2013-2016, CESG15-DE-3114. 

We acknowledge the EuroHPC Joint Undertaking for awarding us access to the supercomputer Leonardo, hosted by CINECA (Italy), and to the supercomputer Marenostrum 5, hosted by the Barcelona Supercomputing Center (Spain), through an EuroHPC Benchmark Access call.

\bibliographystyle{JHEP_mod}
\bibliography{bibliography}

\end{document}